# On the Distribution of the Sum of Gamma-Gamma Variates and Applications in RF and Optical Wireless Communications

Nestor D. Chatzidiamantis, *Student Member*, IEEE, and George K. Karagiannidis, *Senior Member, IEEE*

## Abstract

The Gamma-Gamma (GG) distribution has recently attracted the interest within the research community due to its involvement in various communication systems. In the context of RF wireless communications, GG distribution accurately models the power statistics in composite shadowing/fading channels as well as in cascade multipath fading channels, while in optical wireless (OW) systems, it describes the fluctuations of the irradiance of optical signals distorted by atmospheric turbulence. Although GG channel model offers analytical tractability in the analysis of single input single output (SISO) wireless systems, difficulties arise when studying multiple input multiple output (MIMO) systems, where the distribution of the sum of independent GG variates is required. In this paper, we present a novel simple closed-form approximation for the distribution of the sum of independent, but not necessarily identically distributed GG variates. It is shown that the probability density function (PDF) of the GG sum can be efficiently approximated either by the PDF of a single GG distribution, or by a finite weighted sum of PDFs of GG distributions. To reveal the importance of the proposed approximation, the performance of RF wireless systems in the presence of composite fading, as well as MIMO OW systems impaired by atmospheric turbulence, are investigated. Numerical results and simulations illustrate the accuracy of the proposed approach.

The authors are with the Wireless Communications Systems Group (WCSG), Department of Electrical and Computer Engineering, Aristotle University of Thessaloniki, GR-54124 Thessaloniki, Greece (e-mails: {nestoras, geokarag}@auth.gr).

## I. INTRODUCTION

In communication theory, channel statistical modeling is very crucial, since it can be applied in the design and performance evaluation of various communication systems. A distribution which has recently attracted the interest within the research community due to its involvement in various communication systems, is the so-called Gamma-Gamma (GG) distribution. This distribution is equivalent to the squared Generalized-$K$ ($K_G$) distribution [1] and can be derived from the product of two independent Gamma random variables (RVs). Moreover, for certain values of its parameters, it coincides with the $K$-distribution, which in the past has been widely used in a variety of applications, including the statistical characterization of the intensity of SAR images [2], as well as to model the statistics of the reverbation intensity in underwater communications [3]- [4]. Of particular interest is the application of the GG distribution in optical wireless (OW) systems, where transmission of optical signals through the atmosphere is involved. In these systems, a major performance limiting factor is the *turbulence induced fading*, i.e., rapid fluctuations of the irradiance of the propagated optical signals caused by atmospheric turbulence, which can be accurately modeled using the statistics of the GG distribution [5]. Furthermore, in recent years, GG distribution has also been applied in the field of RF wireless communications; specifically, to model the power statistics in composite fading channels [1], [6]. Additionally, since it includes the well known Double-Rayleigh model [7], GG distribution can be further employed to model the power statistics in cascade multipath fading channels (which occur, e.g., in keyhole or in mobile-to-mobile communication scenarios [8]). The reason for employing GG model in RF wireless systems, is that this distribution is general enough to accurately capture the effects of the combined shadowing and multipath fading or cascade multipath fading which are encountered in mobile communication channels.

Although GG channel model is analytically tractable in the performance analysis of various single input single output (SISO) wireless communication systems [1], [9]- [12], difficulties arise when studying the performance of certain diversity schemes in multiple input multiple output (MIMO) wireless systems. Specifically, these difficulties appear when the distribution of the sum of independent GG variates is required and have their origin in the fact that a straight derivation of this distribution is analytically infeasible, due to the involvement of the modified Bessel function of the second kind.

In the past, a limited number of published works dealing with the distribution of the sum of independent GG variates and its application in communications systems, appeared in the technical literature. In the context of mobile communications, where GG distribution models the statistics of the signal-to-noise ratio (SNR) in the presence of $K_G$ fading channel model, the maximal ratio combining (MRC) receiver has been investigated in [13] and [14]. However, in [13], the sum of independent GG variates was not investigated (shadowing term was common in every diversity branch), while in [14] the expressions that were derived for the statistics of the SNR at the output of the combiner, do not hold[1]. In the context of OW communications, the statistics of the sum of GG variates have been used in the performance analysis of MIMO systems operating over GG turbulence model and employing equal gain combining (EGC) at the receiver. In [15], the performance of such a system was investigated over identical OW links, using an infinite power series representation for the probability density function (PDF) of the turbulence-induced fading term at the output of the receiver. Although this approach allowed the derivation of simple and accurate expressions at the high SNR regime, it was not computationally attractive when the number of the transmit/receive apertures increased and/or the underlying OW links were non identically distributed.

In this paper, we address to this cumbersome statistical problem by applying a novel and simpler approach. We present novel closed-form expressions that approximate efficiently the PDF of the distribution of the sum of independent GG variates based on the following well known issues: a) GG distribution is derived from the product of two independently distributed Gamma RVs, and b) the distribution of the sum of Gamma variates is analytically tractable. Our analysis encompasses the case where the variates involved in the sum are identically distributed, as well as the case of non identical variates. Furthermore, in order to reveal the importance of the proposed statistical formulation, we study the performance of two different systems; an RF wireless system operating over the $K_G$ fading model and employing MRC diversity scheme at the receiver and a MIMO OW system operating over strong turbulence channels and employing EGC at the receiver. For these systems, closed-form expressions that approximate significant system performance metrics, such as bit error rate (BER) and outage probability, are obtained.

The remainder of the paper is organized as follows. After a brief introduction of the GG distribution in Section II, novel closed-form expressions that approximate the PDF of the sum of GG variates are obtained in Section III. Moreover, in Section IV the obtained results are applied to derive closed-form expressions for the performance evaluation of an MRC receiver operating over the $K_G$ fading model, while in Section V the same results are further applied to the performance evaluation of MIMO OW systems operating over strong turbulence channels and employing EGC at the receiver. Finally, in Section VI, useful concluding remarks are provided.

---

[1]In [14, Eq. A-2], it is assumed that the parameter $a = k - m$ is not an integer. However this comes in contradiction with the assumption that the parameters $k$ and $m$ in [14, Eq. A-4] are integers, which results to the incorrect expressions of [14, Eq. 11] and [14, Eq. 12].

## II. The GG Distribution

Let $\gamma \geqslant 0$, then the PDF of a three-parameter GG RV, which is derived from the square of a $K_G$ distributed RV, is given by [1]

$$f_\gamma\left(\gamma; k, m, \Omega\right) = \frac{2(km)^{\frac{k+m}{2}} \gamma^{\frac{k+m}{2}-1}}{\Gamma\left(m\right)\Gamma\left(k\right)\Omega^{\frac{k+m}{2}}} K_{k-m}\left[2\left(\frac{km}{\Omega}\gamma\right)^{1/2}\right] \tag{1}$$

where $k \geqslant 0$ and $m \geqslant 0$ are the distribution shaping parameters, $K_\nu\left(\cdot\right)$ is the modified Bessel function of order $\nu$ [16, 8.407/1], $\Gamma\left(\cdot\right)$ is the Gamma function [16, 8.310/1] and $\Omega$ is related with the mean as $\mathbf{E}\left[\gamma\right] = \Omega$, with $\mathbf{E}\left[\cdot\right]$ denoting expectation.

The distribution in (1) is generic, since it describes various models frequently used in communication systems, for several combinations of $k$ and $m$. Hence, as $k \to \infty$, it approximates the well-known Gamma distribution (or equivalently squared Nakagami-$m$ [17]), while for $m = 1$, it coincides with the statistics of a squared $K$-distributed[2] RV with PDF given by

$$f_\gamma\left(\gamma; k, \Omega\right) = \frac{2k^{\frac{k+1}{2}} \gamma^{\frac{k-1}{2}}}{\Gamma\left(k\right)\Omega^{\frac{k+1}{2}}} K_{k-m}\left[2\left(\frac{k}{\Omega}\gamma\right)^{1/2}\right]. \tag{2}$$

Furthermore for the special case of $k = 1$ and $m = 1$, it reduces to the power statistics of the double Rayleigh model, frequently used in cascade multipath fading channels, with PDF given by [7]- [8] as

$$f_\gamma\left(\gamma; \Omega\right) = \frac{2}{\Omega} K_0\left[2\sqrt{\frac{\gamma}{\Omega}}\right]. \tag{3}$$

The $n$-th moment of $\gamma$ is given as [1]

$$\mathbf{E}\left[\gamma^n\right] = \xi^{-n} \frac{\Gamma\left(k+n\right)\Gamma\left(m+n\right)}{\Gamma\left(k\right)\Gamma\left(m\right)} \tag{4}$$

where $\xi = \frac{km}{\Omega}$. Moreover its cumulative density function (CDF) has been expressed using [10, Eq. 7] and [16, Eq. 9.31/5] as

$$F_\gamma\left(\gamma; k, m, \Omega\right) = \frac{1}{\Gamma\left(k\right)\Gamma\left(m\right)} G_{1,3}^{2,1}\left[\xi\gamma \left|\begin{array}{c} 1 \\ k, m, 0 \end{array}\right.\right], \tag{5}$$

where $G\left[\cdot\right]$ is Meijer's G function [16, Eq. 9.301].

It is important to note that the GG distribution can be derived from the product of two independent RVs, $x$ and $y$ as [18]

$$\gamma = xy \tag{6}$$

---

[2]In [4] and [5], the statistics of the square of a $K$-distributed RV are also referred as $K$-distribution.

when are both Gamma distributed with PDF given by

$$f_i\left(i; m_i, \eta_i\right) = \frac{i^{m_i - 1}}{\eta_i^{m_i} \Gamma\left(m_i\right)} \exp\left(-\frac{i}{\eta_i}\right), \ i = x, y \tag{7}$$

and parameters $\left(m_x = k, \eta_x = 1/k\right)$ and $\left(m_y = m, \eta_y = \Omega/m\right)$ respectively[3].

## III. Efficient Approximation to the Sum of GG variates

Let us consider $L$ independent GG variates denoted by $\{\gamma_l\}_{l=1}^L$, each having shaping parameters $k_l$ and $m_l$, and mean $\Omega_l$. The sum of $L$ GG variates, $S_\gamma$, is defined as

$$S_\gamma \triangleq \sum_{l=1}^L \gamma_l = \sum_{l=1}^L x_l y_l, \tag{8}$$

where $x_l$ and $y_l$ are Gamma RVs with parameters $\left(k_l, 1/k_l\right)$ and $\left(m_l, \Omega_l/m_l\right)$ respectively. Eq. (8) can be rewritten as

$$S_\gamma = \frac{\left(\sum\limits_{l=1}^L x_l\right)\left(\sum\limits_{l=1}^L y_l\right)}{L} + \frac{1}{L} \sum_{i=1}^{L-1} \sum_{j=i+1}^L \left(x_i - x_j\right)\left(y_i - y_j\right). \tag{9}$$

### A. Identical Variates

When the variates of the sum in (8) are independent and identically distributed (i.i.d.) (i.e. $k_l = k$, $m_l = m$, $\Omega_l = \Omega$), $\{x_l\}_{l=1}^L$ and $\{y_l\}_{l=1}^L$ are also identically distributed. Hence, according to (9), the unknown distribution of $S_\gamma$ can be approximated by the distribution of the RV $\hat{S}_\gamma$, which is defined as

$$S_\gamma \approx \hat{S}_\gamma = \frac{\left(\sum\limits_{l=1}^L x_l\right)\left(\sum\limits_{l=1}^L y_l\right)}{L}, \tag{10}$$

with the approximation error, $\varepsilon$, given by

$$\varepsilon = \frac{1}{L} \sum_{i=1}^{L-1} \sum_{j=i+1}^L \left(x_i - x_j\right)\left(y_i - y_j\right). \tag{11}$$

Equivalently, (10) can be written as the product of two RVs $s_1$ and $s_2$, i.e.

$$\hat{S}_\gamma = s_1 s_2, \tag{12}$$

where

$$s_1 = \frac{1}{L} \sum_{l=1}^L x_l \tag{13}$$

---

[3]It follows from symmetry that it is equivalent to consider the set of parameters $\left(m_x = k, \eta_x = \frac{\Omega}{k}\right)$ and $\left(m_y = m, \eta_y = \frac{1}{m}\right)$.

and

$$s_2 = \sum_{l=1}^{L} y_l. \tag{14}$$

Since the sum of i.i.d. Gamma variates remains Gamma distributed [19], it can be proved that $s_1$ and $s_2$ are both Gamma distributed with set of parameters $\left(Lk, \frac{1}{Lk}\right)$ and $\left(Lm, \frac{L\Omega}{Lm}\right)$ respectively. Hence, according to (6), $\hat{S}_\gamma$ will be GG distributed with shaping parameters

$$k_{\hat{S}_\gamma} = Lk, \tag{15}$$

$$m_{\hat{S}_\gamma} = Lm, \tag{16}$$

and mean

$$\Omega_{\hat{S}_\gamma} = L\Omega. \tag{17}$$

The accuracy of the proposed approximation depends on the approximation error defined in (11). The exact PDF of the error is difficult to be derived due to its complicate definition[4]. However, its first moments, which are also indicatives of its statistical behaviour, can be calculated using (11). Specifically, according to the Appendix, the mean of $\varepsilon$ is equal to 0, while its variance depends from $k$, $m$, $\Omega$, and $L$, according to

$$\mathbf{E}\left[\varepsilon^2\right] = (L-1)\frac{\Omega^2}{km}. \tag{18}$$

It is obvious from the above equation that the variance of the approximation error increases for a certain combination of $k$, $m$ and $\Omega$, as the number of the RVs of the sum in (8) increases, having as a result the approximating distribution of $\hat{S}_\gamma$ to loose its accuracy.

In order to improve the accuracy of the proposed approximation, an adjustment parameter is introduced that modifies the shaping parameters of the approximating distribution of $\hat{S}_\gamma$. Specifically, we assume that the maximum of the shaping parameters[5] $k_{\hat{S}_\gamma}$ is modified by an adjustment parameter, $\varepsilon_\gamma$, according to

$$k_{\hat{S}_\gamma} = Lk + \varepsilon_\gamma. \tag{19}$$

The adjustment parameter, $\varepsilon_\gamma$, is evaluated through the following optimization problem

$$\varepsilon_\gamma = \arg\min_{\varepsilon_\gamma} \left| \mathbf{E}\left[\hat{S}_\gamma^\nu\right] - \mathbf{E}\left[S_\gamma^\nu\right] \right|, \nu = 1, ...4 \tag{20}$$

---

[4]It was observed that it converges to the Laplacian distribution as the number of the RVs of the sum in (8) increases.

[5]Since $K_{-\nu}(x) = K_\nu(x)$, we assume for convenience and without loss of generality that $k_{\hat{S}_\gamma} \geqslant m_{\hat{S}_\gamma}$.

where $\mathbf{E}\left[\hat{S}_\gamma^\nu\right]$ are the moments of the distribution of $\hat{S}_\gamma$, which can be derived from (4) using the parameters $\left(k_{\hat{S}_\gamma}, m_{\hat{S}_\gamma}, \Omega_{\hat{S}_\gamma}\right)$, as provided by (19), (16) and (17) respectively; $\mathbf{E}\left[S_\gamma^\nu\right]$ are the moments of the distribution of $S_\gamma$, calculated using the multinomial expansion, according to

$$\mathbf{E}\left[S_\gamma^\nu\right] = \sum_{\nu_1=0}^\nu \sum_{\nu_2=0}^{\nu_1} ... \sum_{\nu_{L-1}=0}^{\nu_{L-2}} \binom{\nu}{\nu_1}\binom{\nu_1}{\nu_2}...\binom{\nu_{L-2}}{\nu_{L-1}} \mathbf{E}\left[\gamma_1^{\nu-\nu_1}\right] \mathbf{E}\left[\gamma_2^{\nu_1-\nu_2}\right] ... \mathbf{E}\left[\gamma_L^{\nu_{L-1}}\right] \qquad (21)$$

and using (4) for the respective parameters $(k, m, \Omega)$.

The optimization problem in (20) is a nonlinear multiple function problem, which is difficult to solve analytically, yet not impossible to derive an approximative solution numerically. After applying non-linear regression methods [20], it was found that the adjustment parameter depends on $L$, $k$ and $m$ with a function of the form of

$$\varepsilon_\gamma\left(L, k, m\right) = (L-1)\frac{-0.127 - 0.95k - 0.0058m}{1 + 0.00124k + 0.98m}. \qquad (22)$$

Hence, using (22) in conjunction with (19), (16) and (17), the parameters of a single GG distribution are defined, which accurately approximates the distribution of the sum of $L$ i.i.d. GG variates.

### B. Non-Identical Variates

When the GG variates of the sum in (8) are independent and not identically distributed (i.n.i.d.), but have one shaping parameter in common, as it happens in most practical applications, (i.e. $k_l = k$, but $m_l$ and $\Omega_l$ are different), the unknown PDF of the sum can still be approximated by the PDF of the RV $\hat{S}_\gamma$, as defined by (12).

As in the i.i.d. case, $\hat{S}_\gamma$ can be written as the product of two RVs $s_1$ and $s_2$, defined by (13) and (14) respectively. Since $k_l = k$, $\{x_l\}_{l=1}^L$ are i.i.d. and $s_1$ is Gamma distributed with parameters $\left(Lk, \frac{1}{Lk}\right)$. However, the derivation of the distribution of $s_2$ is not straightforward, since $\{y_l\}_{l=1}^L$ are not identically distributed. In order to derive the PDF of $s_2$, the exact closed-form expressions for the sum of non-identical Gamma variates presented in [21] are used. According to this approach, the PDF of $s_2$ can be written as a nested finite weighted sum of Gamma PDFs,

$$f_{s_2}\left(z\right) = \sum_{i=1}^L \sum_{j=1}^{m_i} w_L\left(i, j, \{m_l\}_{l=1}^L, \{\Omega_l\}_{l=1}^L\right) f_y\left(z; j, \frac{\Omega_i}{m_i}\right), \qquad (23)$$

where $y$ is a Gamma distributed RV with PDF defined by (7) and the weights can be easily and quickly evaluated using the recursive formula of [21, Eq. (8)] for the parameters $\{m_l\}_{l=1}^L$ and $\{\Omega_l\}_{l=1}^L$, according

to

$$w_L\left(i, m_i - t, \{m_l\}_{l=1}^L, \{\Omega_l\}_{l=1}^L\right) = \frac{1}{t}\sum_{\substack{q=1 \\ q\neq i}}^L \sum_{j=1}^t \frac{m_q}{\frac{\Omega_i^j}{m_i^j}}\left(\frac{m_i}{\Omega_i} - \frac{m_q}{\Omega_q}\right)^{-j} w_L\left(i, m_i - t + j, \{m_l\}_{l=1}^L, \{\Omega_l\}_{l=1}^L\right)$$

(24)

where $t = 1, ..., m_i - 1$ and with

$$w_L\left(i, m_i, \{m_l\}_{l=1}^L, \{\Omega_l\}_{l=1}^L\right) = \frac{\frac{\Omega_i^{m_i}}{m_i^{m_i}}}{\prod_{h=1}^L \frac{\Omega_h^{m_h}}{m_h^{m_h}}} \prod_{\substack{j=1 \\ j\neq i}}^L \left(\frac{m_j}{\Omega_j} - \frac{m_i}{\Omega_i}\right)^{-m_j}.$$

(25)

The PDF of the product of $s_1$ and $s_2$ is evaluated as

$$f_{\hat{S}_\gamma}(z) = \int_0^\infty \frac{1}{x} f_{s_1}\left(x; Lk, \frac{1}{Lk}\right) f_{s_2}\left(\frac{z}{x}\right) dx.$$

(26)

Using (23) and [16, Eq. 3.471/9], Eq. (26) yields as

$$f_{\hat{S}_\gamma}(z) = \sum_{i=1}^L \sum_{j=1}^{m_i} w_L\left(i, j, \{m_l\}_{l=1}^L, \{\Omega_l\}_{l=1}^L\right) f_\gamma\left(z; Lk, j, \frac{j\Omega_i}{m_i}\right),$$

(27)

where $\gamma$ is a GG distributed RV with PDF defined by (1). Hence, an efficient approximation to the PDF of the sum of $L$ non-identical GG variates, when one of the shaping parameters remains the same for all variates[6], can be a nested finite weighted sum of GG PDFs.

## IV. Application in RF Wireless Systems

### A. System Model

Let us consider a diversity receiver with $L$ branches operating over the composite $K_G$ fading channel [14]. The equivalent complex baseband received signal at the $l$th ($l = 1, 2, ..., L$) branch is given by

$$z_l = sh_l + n_l$$

(28)

where $s$ is the transmitted complex symbol with energy $\mathbf{E}\left[\left|s^2\right|\right]$ and $|\cdot|$ denoting absolute value, $h_l$ is the channel's complex gain in the path between the transmitter and the $l$th branch, and $n_l$ is the complex Additive White Gaussian Noise (AWGN), having single sided power spectral density $N_o$ and assumed to be identical in all branches.

---

[6]Note that due to symmetry, the same approximation also holds when $m_i = m$ and $k_i$ are different, by interchanging $k_i$ and $m_i$ in (27), i.e. $m_i = k_i$ and $k = m$.

Since operation over $\mathrm{K}_G$ fading channel model is considered, the square of the fading envelope, $R_l^2 = |h_l|^2$, is statistically described by the PDF of Eq. (1) with shaping parameters $k_l$ and $m_l$, and mean $\mathbf{E}\left[R_l^2\right] = \Omega_l$. It follows that the instantaneous SNR of the $l$th receiving branch, which is defined as

$$\gamma_l = \frac{R_l^2 E_s}{N_o}, \tag{29}$$

is also GG distributed with PDF given by (1), shaping parameters equal to $k_l$ and $m_l$, and mean equal to the average input SNR of the branch defined by

$$\overline{\gamma}_l = \frac{\Omega_l E_s}{N_o}. \tag{30}$$

Furthermore, maximum ratio combining (MRC) technique is applied at the receiver and hence the total SNR per symbol at the output of the receiver is

$$\gamma_T = \sum_{l=1}^{L} \gamma_l = \frac{E_s}{N_o} \sum_{l=1}^{L} R_l^2. \tag{31}$$

In the analysis that follows, it is assumed that both shadowing and multipath fading effects are independent among the diversity branches, i.e. macrodiversity is studied. As a consequence, the diversity branches are considered independent. Moreover, since shadowing occurs in large geographical areas, it is further assumed that the parameter that statistically describes the channel's shadowing effects, i.e. $k_l$, remains constant among the diversity branches, i.e. $k_l = k$ [14]. Note that the assumption of the uncorrelated diversity branches can be further applied in the scenario of diversity reception in cascade multipath fading channels, when a rich scattering radio environment is considered (see [8] for examples).

## B. Error Analysis

The average BER of the under consideration RF system can be evaluated directly by averaging the conditional BER, $P_e\left(\gamma\right)$, which depends from the type of modulation, over the PDF of $\gamma_T$, $f_{\gamma_T}\left(\gamma\right)$, i.e.

$$\overline{P}_{be} = \int_0^\infty P_e\left(\gamma\right) f_{\gamma_T}\left(\gamma\right) d\gamma. \tag{32}$$

Without loss of generality, two types of modulation are considered, binary phase shift keying (BPSK) and differential binary phase shift keying (DBPSK). For BPSK and for high values of the average input SNR, the conditional BER is given by $P_e\left(\gamma\right) = \frac{1}{2}\mathrm{erfc}\left(\sqrt{\gamma}\right)$, where $\mathrm{erfc}\left(\cdot\right)$ is the complimentary error function [16, Eq. (8.250/1)], while for DBPSK is given by $P_e\left(\gamma\right) = \frac{1}{2}\exp\left(-\gamma\right)$ [22]. Hence, the average BER of the diversity system can be evaluated by

$$\overline{P}_{be} = \frac{1}{2} \int_0^\infty \mathrm{erfc}\left(\sqrt{\gamma}\right) f_{\gamma_T}\left(\gamma\right) d\gamma \tag{33}$$

 

and

$$\overline{P}_{be} = \frac{1}{2} \int_0^\infty \exp\left(-\gamma\right) f_{\gamma_T}\left(\gamma\right) d\gamma \tag{34}$$

when BPSK and DBPSK modulation schemes are employed respectively.

*1) Independent and Identically Distributed Diversity Branches:* When the received signals at the diversity branches are independent and identically distributed (i.i.d.), i.e. $m_l = m$ and $\overline{\gamma}_l = \overline{\gamma}$, the approximation for the PDF of the sum of i.i.d. GG variates can be used in order to evaluate the integrals in (33) and (34). Specifically, the PDF of $\gamma_T$ can be approximated by the PDF of a single GG variate with parameters defined by (19), (16) and (17), i.e.

$$f_{\gamma_T}\left(\gamma\right) \approx f_\gamma\left(\gamma; k_T, m_T, \overline{\gamma}_T\right) \tag{35}$$

where $k_T = Lk + \varepsilon_{\gamma_1}$, $m_T = Lm$, $\overline{\gamma}_T = L\overline{\gamma}$ and $\varepsilon_{\gamma_1}$ is defined from (22) for the set of parameters of $(L, k, m)$. By substituting (35) in (33) and (34), and using the corresponding BER expressions of a Single Input Single Output (SISO) system with parameters $(k_T, m_T, \overline{\gamma}_T)$, analytical expressions that approximate the average BER of the diversity system can be derived. Hence, using [1, Eq. (8)], the average BER for BPSK modulation can be approximated by

$$\overline{P}_{be} \approx \frac{\xi_T^{\frac{k_T+m_T}{2}}}{2\sqrt{\pi}\Gamma\left(k_T\right)\Gamma\left(m_T\right)} G_{2,3}^{2,2}\left[\xi_T \left|\begin{array}{c} \frac{1-\beta_T}{2}, -\frac{\beta_T}{2} \\ \frac{a_T}{2}, -\frac{a_T}{2}, -\frac{\beta_T+1}{2} \end{array}\right.\right], \tag{36}$$

where $\xi_T = \frac{k_T m_T}{\overline{\gamma}_T}$, $\beta_T = k_T + m_T - 1$ and $a_T = k_T - m_T$. In a similar manner, using [1, Eq. (9)], the BER performance of the DBPSK diversity system is evaluated as

$$\overline{P}_{be} \approx \frac{1}{2}\xi_T^{\frac{\beta_T}{2}} \exp\left(\frac{\xi_T}{2}\right) W_{-\frac{\beta_T}{2}, \frac{a_T}{2}}\left(\xi_T\right), \tag{37}$$

where $W_{\lambda,\mu}\left(\cdot\right)$ is the Whittaker function [16, Eq. 9.220].

*2) Independent, but not Necessarily Identically Distributed Diversity Branches:* When the received signals at the diversity branches are independent, but not necessarily identically distributed (i.n.i.d.), i.e. $m_l$ and $\overline{\gamma}_l$ are different among the diversity branches, the approximation for the PDF of the sum of i.n.i.d. GG variates can be used in order to approximate the average BER of the under consideration diversity system. Hence, according to (27), the PDF of $\gamma_T$ can be approximated by a nested finite weighted sum of GG PDFs, i.e.

$$f_{\gamma_T}\left(\gamma\right) \approx \sum_{i=1}^{L} \sum_{j=1}^{m_i} \Xi\left(i, j\right) f_\gamma\left(\gamma; Lk, j, \frac{j\overline{\gamma}_i}{m_i}\right), \tag{38}$$

where

$$\Xi\left(i, j\right) = w_L\left(i, j, \{m_l\}_{l=1}^{L}, \{\overline{\gamma}_l\}_{l=1}^{L}\right) \tag{39}$$

are evaluated using (24) and (25). By substituting the above sum in (33) and (34), the average BER of the system for BPSK modulation is approximated by

$$\overline{P}_{be} \approx \sum_{i=1}^{L} \sum_{j=1}^{m_i} \Xi(i,j) \frac{1}{2} \int_0^{\infty} \operatorname{erfc}(\sqrt{\gamma}) f_\gamma\left(\gamma; Lk, j, \frac{j\overline{\gamma}_i}{m_i}\right) d\gamma \tag{40}$$

while for DPSK modulation, by

$$\overline{P}_{be} \approx \sum_{i=1}^{L} \sum_{j=1}^{m_i} \Xi(i,j) \frac{1}{2} \int_0^{\infty} \exp(-\gamma) f_\gamma\left(\gamma; Lk, j, \frac{j\overline{\gamma}_i}{m_i}\right) d\gamma. \tag{41}$$

The integrals in the above equations can be evaluated using the corresponding BER expressions of a SISO system with parameters $\left(Lk, j, \frac{j\overline{\gamma}_i}{m_i}\right)$. Hence, using [1, Eq. (8)], the analytical expression that approximates the average BER of the BPSK diversity system yields as

$$\overline{P}_{be} \approx \sum_{i=1}^{L} \sum_{j=1}^{m_i} \frac{\Xi(i,j)\, \xi_i^{\frac{Lk+j}{2}}}{2\sqrt{\pi}\Gamma(Lk)\,\Gamma(j)} G_{2,3}^{2,2}\left[\xi_i \left| \begin{array}{c} \frac{1-\beta_j}{2}, -\frac{\beta_j}{2} \\ a_j, -a_j, -\frac{\beta_j+1}{2} \end{array} \right. \right], \tag{42}$$

where $\xi_i = \frac{Lkm_i}{\overline{\gamma}_i}$, $a_j = \frac{Lk-j}{2}$ and $\beta_j = Lk + j - 1$. Moreover, using [1, Eq. (9)], the average BER of the DBPSK diversity system is derived as

$$\overline{P}_{be} \approx \sum_{i=1}^{L} \sum_{j=1}^{m_i} \frac{\Xi(i,j)}{2} \xi_i^{\frac{Lk+j-1}{2}} \exp\left(\frac{\xi_i}{2}\right) W_{-\frac{\beta_j}{2}, \frac{a_j}{2}}(\xi_i). \tag{43}$$

### C. Outage Probability

*Outage probability* is defined as the probability that the output SNR of the under consideration diversity system falls below a specified threshold $\gamma_{th}$, which represents a protection value of the SNR above which the quality of the channel is satisfactory. Hence, outage probability can be evaluated by

$$P_{out} = \Pr(\gamma_T \le \gamma_{th}) = \int_0^{\gamma_{th}} f_{\gamma_T}(\gamma)\, d\gamma. \tag{44}$$

*1) Independent and Identically Distributed Diversity Branches:* When the received signals at the diversity branches are i.i.d., i.e. $m_l = m$ and $\overline{\gamma}_l = \overline{\gamma}$, the PDF of $\gamma_T$ can be approximated by a single GG distributed variate with parameters $(k_T, m_T, \overline{\gamma}_T)$ according to (35). Hence, the outage probability of the diversity system can be approximated by

$$P_{out} \approx \int_0^{\gamma_{th}} f_\gamma(\gamma; k_T, m_T, \overline{\gamma}_T)\, d\gamma, \tag{45}$$

which is equivalent to the outage probability of a SISO system with parameters $(k_T, m_T, \overline{\gamma}_T)$. Using (5), (45) yields

$$P_{out} \approx \frac{1}{\Gamma(k_T)\,\Gamma(m_T)} G_{1,3}^{2,1}\left[\xi_T \gamma_{th} \left| \begin{array}{c} 1 \\ k_T, m_T, 0 \end{array} \right. \right] \tag{46}$$

where $\xi_T = \frac{k_T m_T}{\overline{\gamma}_T}$.

*2) Independent but not Necessarily Identically Distributed Diversity Branches:* When the received signals at the diversity branches are i.n.i.d., i.e. $m_l$ and $\overline{\gamma}_l$ are different among the diversity branches, the PDF of $\gamma_T$ is approximated by a nested finite weighted sum of GG PDFs, according to (38). Hence, the outage probability can be approximated by

$$P_{out} \approx \sum_{i=1}^{L} \sum_{j=1}^{m_i} \Xi\left(i, j\right) \int_{0}^{\gamma_{th}} f_{\gamma}\left(\gamma; Lk, j, \frac{j\overline{\gamma}_i}{m_i}\right) d\gamma \tag{47}$$

which is equivalent to a nested finite weighted sum of outage probabilities of SISO systems, each having the parameters $\left(Lk, j, \frac{j\overline{\gamma}_i}{m_i}\right)$. Using (5), (47) yields

$$P_{out} \approx \sum_{i=1}^{L} \sum_{j=1}^{m_i} \Xi\left(i, j\right) \frac{1}{\Gamma\left(Lk\right)\Gamma\left(j\right)} G_{1,3}^{2,1}\left[\xi_i \gamma_{th} \left| \begin{array}{c} 1 \\ Lk, j, 0 \end{array} \right. \right] \tag{48}$$

where $\xi_i = \frac{Lkm_i}{\overline{\gamma}_i}$.

## D. Numerical Results and Discussion

Figs. 1 and 2 illustrate the average BER and outage probability of MRC receivers operating over the $K_G$ fading model, when the diversity branches have the same shaping parameters and average branch input SNR, i.e. $k_l = k$, $m_l = m$ and $\overline{\gamma}_l = \overline{\gamma}$. Approximative analytical results, using Eq. (36) and (37) for the BER evaluation and (46) for the outage probability evaluation, are plotted in comparison with Monte-Carlo (MC) simulation results, for an arbitrary number of diversity branches and assuming a certain combination of shaping parameters ($k = 2$ and $m = 5$). It is observed that there is an excellent match between simulation and the approximative results for every input SNR and normalized outage, $\gamma_{th}/\overline{\gamma}$, in both performance metrics. It is also clearly depicted that the approximative analytical expressions remain accurate, even when the number of the diversity branches increases.

In Figs. 3-6, the BER and outage probability performance metrics of MRC receivers are depicted, when the diversity branches are i.n.i.d., i.e. $m_l$ and $\overline{\gamma}_l$ are different among the diversity branches. Specifically, it is assumed that the average input SNR of $l$th branch is given by $\overline{\gamma}_l = \overline{\gamma}_1 \exp\left[-\delta\left(l-1\right)\right]$, where $\bar{\gamma}_1$ is the average input SNR of the first branch and $\delta$ is a decaying factor. Using (40) for BPSK and (41) for DBPSK modulation, Figs. 3-4 present the approximative analytical results for the average BER, as a function of the first brach average input SNR, for several combinations of shaping parameters $k$, $m_l$ and decaying factors $\delta$. It is obvious from the figures that the approximative analytical expressions for the average BER are close to simulation results (their difference is not greater than 3 dB at target BERs equal to $10^{-5}$), even when $\delta$ increases. Moreover it is evident that the proposed approximation acts as a lower bound for high values of SNR and the smaller the number of diversity branches and/or $\delta$ are, the more accurate is the bound. Similar behaviour has been also observed for the outage probability, which is depicted in Figs. 5 and 6 as a function of the first branch normalized outage threshold, $\gamma_{th}/\overline{\gamma}_1$, for the same combinations of $k$, $m_l$ and $\delta$. The difference between analytical results, derived from (48), and simulation results also lies within 3dB in all cases examined, and the proposed approximation acts as a tight lower bound.

## V. Application in Optical Wireless Systems

### A. System Model

Consider a Multiple Input Multiple Output (MIMO) optical wireless (OW) system where the information signal is transmitted via $M$ apertures and received by $N$ apertures over strong atmospheric turbulence conditions. For the OW system under consideration, it is assumed that the information bits are modulated using On-Off keying (OOK) and transmitted through the $M$ apertures using repetition coding [23]. Moreover, a large field of view is considered for each receiver indicating that multiple transmitters are simultaneously observed by each receiver. This actually leads to the collection of larger amount of background radiation which justifies the use of the AWGN model as a good approximation of the Poisson photon counting detection model [24]. Hence, the received signal at the $q$th receive aperture is given by

$$r_q = x\eta \sum_{p=1}^{M} I_{pq} + v_q, \ q = 1,...N \tag{49}$$

where $x \in \{0, 1\}$ represents the information bits, $\eta$ is the optical-to-electrical conversion coefficient and $v_q$ is the AWGN with zero mean and variance $\sigma_v^2 = N_o/2$.

The term $I_{pq}$ denotes the fading coefficient that models the atmospheric turbulence through the optical channel between the $p$th transmit and the $q$th receive aperture. Since operation under strong atmospheric turbulence conditions is assumed, according to [9], the parameter which represents the effective number of small scale scatterers can be considered equal to 1. Hence, the optical channel in the $p$-$q$th transmit-receive pair can be statistically described by a GG distribution with parameters $k = 1$, $m = a_{pq}$ and $\Omega = \mathbf{E}\left[I_{pq}\right]$ [9], where $a_{pq}$ is related to the effective number of large scale scatterers. Furthermore, it is assumed that the statistics of the fading coefficients of the underlying channels are statistically independent; an assumption which is realistic by placing the transmitter and the receiver apertures just a few centimeters apart [25].

At the receiver side, the received optical signals from the $N$ apertures are combined using equal gain combining (EGC), which is an efficient combing scheme in OW systems [24]- [25]. Hence, the output

of the receiver is

$$r = \sum_{q=1}^{N} r_q = \frac{x\eta}{MN} \sum_{q=1}^{N} \sum_{p=1}^{M} I_{pq} + \upsilon. \tag{50}$$

Note that a scaling factor of $MN$ appears in (50). The factor $M$ is included in order to ensure that the total transmit power is the same with that of a system with no transmit diversity, while the factor $N$ ensures that the sum of the $N$ receive aperture areas is the same with the aperture area of a system with no receive diversity.

The received electrical SNR of the OW link between the $p$ transmit and $q$ receive aperture, can be defined as [26]

$$h_{pq} = \frac{\eta^2 I_{pq}^2}{N_o}, \tag{51}$$

while its average as $\mu_{pq} = \frac{\eta^2 E[I_{pq}]^2}{N_o}$. According to the above definitions, the electrical SNR of the combined signal at the output of the receiver, becomes

$$h_T = \frac{\eta^2 (I_T)^2}{M^2 N^2 N_o}, \tag{52}$$

where $I_T = \sum_{q=1}^{N} \sum_{p=1}^{M} I_{pq}$.

### B. Error Analysis

The BER probability of the MIMO OW system under consideration, assuming perfect Channel State Information (CSI), is given by [24] as

$$P_e = \int_{\boldsymbol{I}} f_{\boldsymbol{I}} (\boldsymbol{I}) Q \left( \frac{\eta}{2MN\sigma_\upsilon} \sum_{q=1}^{N} \sum_{p=1}^{M} I_{pq} \right) \mathrm{d}\boldsymbol{I} \tag{53}$$

where $f_{\boldsymbol{I}} (\boldsymbol{I})$ is the joint PDF of the vector $\boldsymbol{I} = (I_{11}, I_{12}, ... I_{MN})$ of length $MN$. Furthermore, $Q (\cdot)$ is the Gaussian-Q function defined as $Q (y) = \left( 1/\sqrt{2\pi} \right) \int_y^\infty \exp \left( -t^2/2 \right) dt$ and related to $\mathrm{erfc} (\cdot)$ by $\mathrm{erfc} (x) = 2Q \left( \sqrt{2}x \right)$. Equivalently, Eq. (53) can be evaluated as

$$P_e = \frac{1}{2} \int_0^\infty f_{I_T} (I) \, \mathrm{erfc} \left( \frac{\eta}{2\sqrt{2}\mathrm{NM}\sigma_\upsilon} \mathrm{I} \right) \mathrm{dI} \tag{54}$$

where $f_{I_T} (I)$ is the PDF of $I_T$.

*1) Independent and Identically Distributed OW Links:* When the turbulence induced fading coefficients of the underlying optical links of the MIMO system are independent and identically distributed, i.e. $a_{pq} = a$ and $\mathbf{E} [I_{pq}] = I_o$, the PDF of $I_T$ can be approximated by the PDF of a single GG variate, i.e.

$$f_{I_T} (I) \approx f_\gamma (I; k_T, m_T, \Omega_T) \tag{55}$$

where $k_T = MNa + \varepsilon_{\gamma_2}$, $m_T = MN$, $\Omega_T = MNI_o$ and $\varepsilon_{\gamma_2}$ is calculated from (22) for the parameters $(MN, a, 1)$. Hence, the BER probability of (54) is approximated by

$$P_e \approx \frac{1}{2} \int_0^\infty f_\gamma \left( I; k_T, m_T, \Omega_T \right) \operatorname{erfc} \left( \frac{\eta}{2\sqrt{2} \mathrm{NM} \sigma_v} \mathrm{I} \right) \mathrm{dI}. \tag{56}$$

The integral of (56) can be solved using Meijer's G-functions and their properties. Hence, by substituting the PDF of the GG distribution according to (1), expressing the $K_\nu(\cdot)$ and the $\operatorname{erfc}(\cdot)$ integrands in terms of Meijer's G-function according to [27, Eq. (8.4.23.1)] and [27, Eq. (8.4.14.2)] respectively, and using [27, Eq. (2.24.1.1)], the BER is expressed as

$$P_e \approx \frac{2^{k_T + m_T - 3}}{\sqrt{\pi^3} \Gamma\left( k_T \right) \Gamma\left( m_T \right)} G_{5,2}^{2,4} \left[ \left( \frac{2}{k_T m_T} \right)^2 \mu \; \middle| \; \begin{matrix} \frac{1-k_T}{2}, \frac{2-k_T}{2}, \frac{1-m_T}{2}, \frac{2-m_T}{2}, 1 \\ 0, \frac{1}{2} \end{matrix} \right], \tag{57}$$

where $\mu$ denotes the average electrical SNR of each OW link.

*2) Independent and Not Identically Distributed OW Links:* When the turbulence induced fading coefficients of the underlying optical links of the MIMO OW system are independent, but not identically distributed, the PDF of $I_T$ can be approximated by a nested finite weighted sum of GG PDFs, according to (27), i.e.

$$f_{I_T}(I) \approx \sum_{i=1}^{L} \sum_{j=1}^{m_i} \Xi(i,j) f_\gamma \left( I; L, j, \frac{j \Omega_i}{m_i} \right) \tag{58}$$

where $L = MN$ is the number of the underlying OW links, $m_l = a_{pq}$, $k_l = 1$ and $\Omega_l = \mathbf{E}\left[ I_{pq} \right]$, when $p = 1, ...M$, $q = 1, ...N$ and $l = 1, ...MN$. Furthermore,

$$\Xi(i,j) = w_L \left( i, j, \{m_l\}_{l=1}^L, \{\Omega_l\}_{l=1}^L \right) \tag{59}$$

are evaluated using (24) and (25).

By substituting (58) to (54), the BER probability of the MIMO OW system is approximated by

$$P_e \approx \sum_{i=1}^{L} \sum_{j=1}^{m_i} \Xi(i,j) \frac{1}{2} \int_0^\infty f_\gamma \left( I; L, j, \frac{j \Omega_i}{m_i} \right) \operatorname{erfc} \left( \frac{\eta}{2\sqrt{2} \mathrm{NM} \sigma_v} \mathrm{I} \right) \mathrm{dI} \tag{60}$$

The integral in the above equation can be evaluated by expressing its integrands in terms of Meijer's G-function, as in the i.i.d. case. Hence, the probability of error is given by

$$P_e \approx \sum_{i=1}^{L} \sum_{j=1}^{m_i} \frac{2^{L+j-3} \Xi(i,j)}{\sqrt{\pi^3} \Gamma(L) \Gamma(j)} G_{5,2}^{2,4} \left[ \left( \frac{2}{L^2 m_i} \right)^2 \mu_i \; \middle| \; \begin{matrix} \frac{1-L}{2}, \frac{2-L}{2}, \frac{1-j}{2}, \frac{2-j}{2}, 1 \\ 0, \frac{1}{2} \end{matrix} \right] \tag{61}$$

where $\mu_i$ is the average electrical SNR of the $i$th OW link.

### C. Outage Probability

As in wireless RF systems, the outage probability is defined as the probability that the SNR of the combined signal at the output of the receiver, falls below a specified threshold $h_{th}$. It is considered as an important parameter for OW links to be operated as a part of a data network and is critical in the design of both transport and network layer.

Hence, according to (52), the resulting outage probability of the system is given by

$$P_{out} = \Pr\left(h_T \le h_{th}\right) = \Pr\left(I_T \le I_{th}\right) \tag{62}$$

where $I_{th} = \frac{NM}{\eta}\sqrt{h_{th}N_o}$ is the normalized threshold.

*1) Independent and Identically Distributed OW Links:* When the underlying channels of the MIMO OW system are independent and identically distributed, the PDF of $I_T$ can be approximated by Eq. (55). The outage probability of the under consideration system can be approximated by

$$P_{out} \approx \int_0^{I_{th}} f_\gamma\left(I; k_T, m_T, \Omega_T\right) \mathrm{d}I \tag{63}$$

which is equivalent to the outage probability of a SISO system operating over the GG turbulence model with parameters $(k_T, m_T, \Omega_T)$. Hence, using (5), a closed-form expression for the outage probability yields as

$$P_{out} \approx \frac{1}{\Gamma\left(k_T\right)\Gamma\left(m_T\right)} G_{1,3}^{2,1}\left[\frac{k_T m_T}{\Omega_T} I_{th} \left| \begin{array}{c} 1 \\ k_T, m_T, 0 \end{array} \right. \right]. \tag{64}$$

*2) Independent and Not Identically Distributed OW Links:* When the underlying channels of the MIMO OW system are independent, but not identically distributed, the PDF of $I_T$ can be approximated by Eq. (58). Hence, the outage probability of the under consideration system can be approximated by

$$P_{out} \approx \sum_{i=1}^{L}\sum_{j=1}^{m_i} \Xi\left(i,j\right) \int_0^{I_{th}} f_\gamma\left(I; L, j, \frac{j\Omega_i}{m_i}\right) \mathrm{d}I. \tag{65}$$

From the above equation, it is evident that the outage probability of the MIMO OW system can be approximated by a finite nested weighted sum of outage probabilities of SISO OW links operating over the GG turbulence model with parameters $\left(L, j, \frac{j\Omega_i}{m_i}\right)$. Using (5), an analytical expression is derived as

$$P_{out} \approx \sum_{i=1}^{L}\sum_{j=1}^{m_i} \frac{\Xi\left(i,j\right)}{\Gamma\left(L\right)\Gamma\left(j\right)} G_{1,3}^{2,1}\left[\frac{Lm_i}{\Omega_i} I_{th} \left| \begin{array}{c} 1 \\ L, j, 0 \end{array} \right. \right]. \tag{66}$$

*D. Numerical Results*

In Figs. 7 and 8, the BER and outage probability of MIMO OW systems operating over identically distributed strong turbulence channels with parameters $a = 4$ or $a = 10$ and $I_o = 1$, are depicted. Analytical results, using (57) and (64), are illustrated in comparison with MC simulations. It is observed that there is an excellent match between the approximation and the simulations in every SNR regime for both performance metrics. It is also clearly depicted that the derived approximative expressions are accurate for every MIMO deployment investigated, irrespective the number of transmit and/or receive apertures.

Figs. 9 and 10 depict the BER and outage probability of various MIMO deployments of transmit and receive apertures over i.n.i.d. strong turbulence channels, i.e. the underlying OW links have different turbulence parameters and different average electrical SNRs. As it is clearly illustrated, the approximative analytical results for both performance metrics, derived from (61) and (66) respectively, are very close to the MC simulation results. Specifically, for the MIMO deployments investigated and for practical values of average BER and outage probability, the difference between analytical and simulation results is not greater than 2 dB. Moreover, it is observed that the proposed approximation acts as a lower bound, which becomes less accurate as the number of the underlying i.n.i.d. OW links increases. However, taking into consideration that the BER performance metric is difficult or even impossible to be evaluated with numerical techniques as the number of the OW links increases [26], the derived closed-form expressions can be considered as reliable alternatives to time consuming MC simulations.

## VI. CONCLUSIONS

We examined the statistics of the sum of independent and not necessarily identical GG RVs. Novel closed-form expressions were derived that approximated its PDF either with the PDF of a single GG distribution, when all the variates of the sum were identically distributed, or with a finite weighted sum of PDFs of GG distributions, when the variates of the sum were non identically distributed. Based on the obtained statistical formulas, the performance of MRC diversity receivers operating over the $K_G$ fading channel, as well as MIMO OW systems operating over strong turbulence channels and employing EGC at the receiver, was investigated and major performance metrics were analytically evaluated. The comparison between approximative analytical results and simulations demonstrated that the proposed approximation is accurate when the diversity branches or underlying OW links are identical, while it serves as a tight lower bound for non identical diversity branches or OW links.

## Appendix

Since both $\{x_l\}_{l=1}^{l=L}$ and $\{y_l\}_{l=1}^{l=L}$ are identically distributed squared-Nakagami variates with parameters $(k, 1/k)$ and $(m, \Omega/m)$, respectively, their first moments and variances are given by

$$\mathbf{E}\left[x_l\right] = 1 \tag{67}$$

$$\mathbf{E}\left[y_l\right] = \Omega \tag{68}$$

$$\mathbf{E}\left[x_l^2\right] - \mathbf{E}\left[x_l\right]^2 = \frac{1}{k} \tag{69}$$

$$\mathbf{E}\left[y_l^2\right] - \mathbf{E}\left[y_l\right]^2 = \frac{\Omega^2}{m} \tag{70}$$

where $l = 1, ... L$.

Due to their independency, the first moment of the the error $\varepsilon$ can be easily calculated, according to

$$\mathbf{E}\left[\varepsilon\right] = \frac{1}{L}\sum_{i=1}^{L-1}\sum_{j=i+1}^{L}\left(\mathbf{E}\left[x_i\right] - \mathbf{E}\left[x_j\right]\right)\left(\mathbf{E}\left[y_i\right] - \mathbf{E}\left[y_j\right]\right) = 0. \tag{71}$$

Furthermore, the variance of $\varepsilon$ is equal to its second moment and it is derived from

$$\mathbf{E}\left[\varepsilon^2\right] = \frac{1}{L^2}\mathbf{E}\left[\left(\sum_{i=1}^{L-1}\sum_{j=i+1}^{L}\left(x_i - x_j\right)\left(y_i - y_j\right)\right)^2\right]. \tag{72}$$

After expanding the sums in (72) and taking their square, the terms that appear are calculated using (67), (68), (69) and (70). Specifically, the following terms appear,

$$\mathbf{E}\left[\left(x_i^2 - x_j^2\right)\left(x_h^2 - x_g^2\right)\left(y_i^2 - y_j^2\right)\left(y_h^2 - y_g^2\right)\right] = \begin{cases} 4\frac{\Omega^2}{km} & \text{if } i = h \text{ and } j = g \\ 0 & \text{if } i \neq h \text{ and } j \neq g \\ \frac{\Omega^2}{km} & \text{if } i = h \text{ and } j \neq g \end{cases} \tag{73}$$

where $i, j, h, g = 1 ... L$. After some algebra and using (73), (72) is simplified to (18).

## References

[1] P. S. Bithas, N. C. Sagias, P. T. Mathiopoulos, G. K. Karagiannidis, and A. A. Rontogiannis, "On the performance analysis of digital communications over Generalized-K fading channels," *IEEE Commun. Letters*, vol. 10, no. 5, pp. 353–355, May 2006.

[2] S. Chitroub, A. Houacine, and B. Sansal, "Statistical characterization and modeling of SAR images," *Elsevier Signal Processing*, vol. 82, no. 1, Nov. 1999.

[3] E. Jakeman and P. Pusey, "A model for non-Rayleigh sea echo," *IEEE Transactions on Antennas and Propagation*, vol. 24, no. 6, pp. 806–814, Nov. 1976.

[4] D. A. Abraham, "Signal excess in $K$-distributed reverbation," *IEEE Journal of Oceanic Engineering*, vol. 28, no. 3, Jul. 2003.

[5] M. A. Al-Habash, L. C. Andrews, and R. L. Philips, "Mathematical model for the irradiance probability density function of a laser beam propagating through turbulent media," *Optical Engineering*, vol. 40, pp. 1554–1562, Aug. 2001.

[6] P. M. Shankar, "Error rates in generalized shadowed fading channels," *Wireless Personal Communications*, vol. 28, no. 4, pp. 233–238, Feb. 2004.

[7] M. Uysal, "Diversity analysis of space-time coding in cascaded Rayleigh fading channels," *IEEE Com. Letters*, vol. 10, no. 3, pp. 165–167, Mar. 2006.

[8] J. S. Salo, H. M. El-Sallabi, and P. Vainikainen, "Impact of double-Rayleigh fading on system performance," in *Proc. Int. Symp. on Wireless Pervasive Computing (ISWPC)*, Phuket, Thailand, 2006.

[9] L. Andrews, R. L. Philips, and C. Y. Hopen, *Laser Beam Scintillation with Applications*. SPIE Press, 2001.

[10] T. A. Tsiftsis, "Performance of heterodyne wireless optical communication systems over gamma-gamma atmospheric turbulence channels," *IEE Electr. Lett.*, vol. 44, no. 5, Feb. 2008.

[11] A. Laourine, M.-S. Alouini, and S. Affes, "On the capacity of generalized-K fading channels," *IEEE Transactions on Wireless Comm.*, vol. 7, no. 7, pp. 2441–2445, Jul. 2008.

[12] M. Uysal, S. M. Navidpour, and J. Li, "Error rate performance of coded Free-Space optical links over strong turbulence channels," *IEEE Communications Letters*, vol. 8, no. 10, pp. 635–637, Oct. 2004.

[13] P. M. Shankar, "Performance analysis of diversity combining algorithms in shadowed fading channels," *Wireless Personal Communications*, vol. 37, no. 1-2, pp. 61–72, Apr. 2006.

[14] P. S. Bithas, P. T. Mathiopoulos, and S. A. Kotsopoulos, "Diversity reception over Generalized-K ($K_G$) fading models," *IEEE Trans. Wir. Comm.*, vol. 6, no. 12, pp. 4238–4243, Dec. 2007.

[15] E. Bayaki, R. Schober, and R. K. Mallik, "Performance analysis of Free-Space Optical sytems in Gamma-Gamma fading," in *Globecom*, New Orleans, USA, 2008.

[16] I. S. Gradshteyn and I. M. Ryzhik, *Table of Integrals, Series, and Products*, 6th ed. New York: Academic, 2000.

[17] M. Nakagami, "The m-distribution–A general formula of intensity distribution of rapid fading," *Statistical Methods in Radio Wave Propagation*, vol. 40, pp. 757–768, Nov. 1962.

[18] G. K. Karagiannidis, T. H. Tsiftsis, and R. K. Mallik, "Bounds for Multihop Relayed Communications in Nakagami-m Fading," *IEEE Trans. on Comm.*, vol. 54, no. 1, pp. 18–22, Jan. 2006.

[19] E. K. Al-Hussaini and A. A. M. Al-Bassiouni, "Performance of MRC diversity systems for the detection of signals with Nakagami fading," *IEEE Trans. on Comm.*, vol. 33, no. 12, pp. 1315–1319, Dec. 1985.

[20] G. A. Seber and C. J. Wild, *Nonlinear Regression*. New York: Wiley, 1989.

[21] G. K. Karagiannidis, N. C. Sagias, and T. A. Tsiftsis, "Closed-form statistics for the sum of Squared Nakagami-m variates and its applications," *IEEE Trans. on Comm.*, vol. 54, no. 8, pp. 1453–1458, 2006.

[22] S. Sampei, *Applications of digital wireless technologies to Global Wireless Communications*. London: Prentice Hall, 1997.

[23] M. Safari and M. Uysal, "Do we really need space-time coding for free-space optical communication with direct detection?" *IEEE Trans. Wireless Commun.*, vol. 7, no. 11, pp. 4445–4448, Nov. 2008.

[24] S. M. Navidpour and M. Uysal, "BER performance of free-space optical transmission with spatial diversity," *IEEE Trans. Wireless Commun.*, vol. 6, no. 8, pp. 2813–2819, Aug. 2007.

[25] E. Lee and V. Chan, "Part 1: Optical communication over the clear turbulent atmospheric channel using diversity," *IEEE Journal on Selected Areas in Commun.*, vol. 22, no. 9, pp. 71 896–1906, Nov. 2004.

[26] T. A. Tsiftsis, H. G. Sandalidis, G. K. Karagiannidis, and M. Uysal, "Optical wireless links with spatial diversity over strong atmospheric turbulence channels," *IEEE Trans. on Wireless Communications*, vol. 8, no. 2, pp. 951–957, Feb. 2009.

[27] A. P. Prudnikov, Y. A. Brychkov, and O. I. Marichev, *Integral and Series. Vol. 3: More Special Functions.* Amsterdam: Gordon and Breach Science Publishers, 1986.

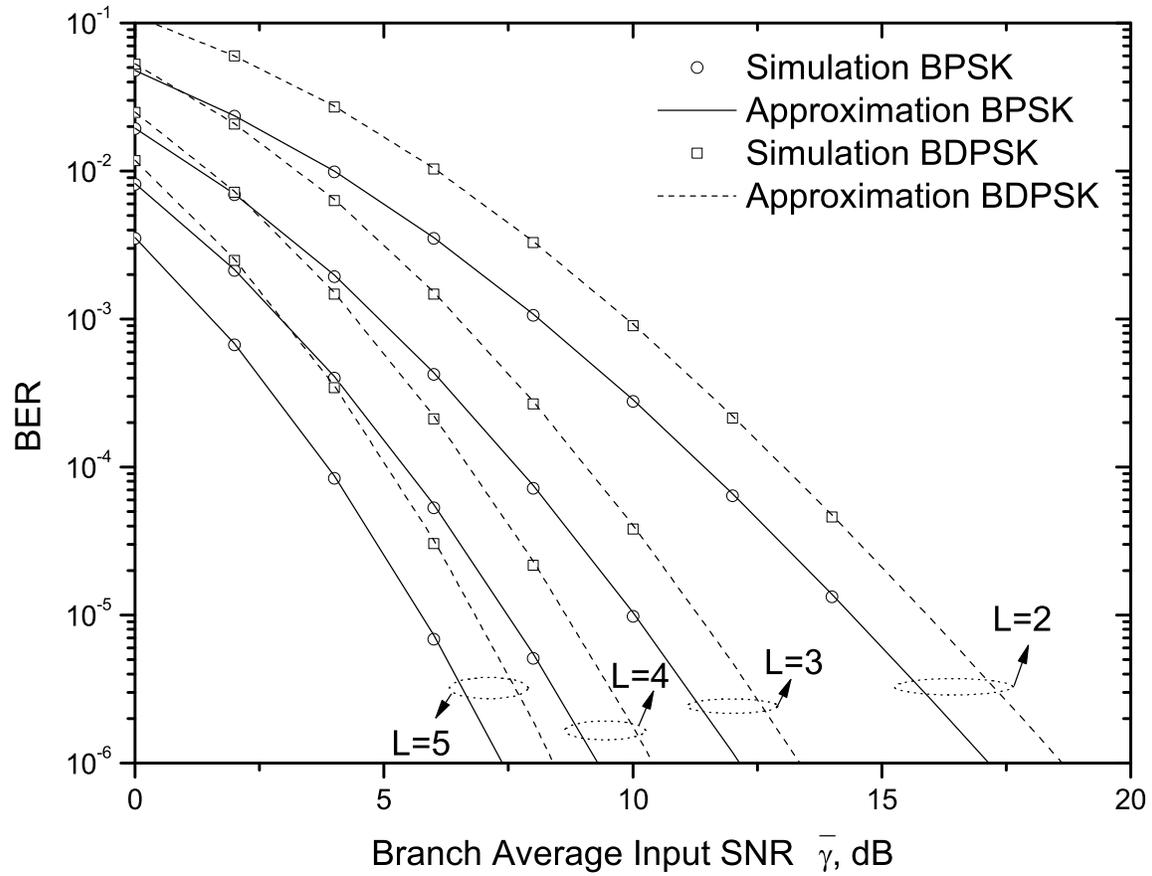

Fig. 1.    Comparison of approximate average BER and MC simulation results of MRC receivers with i.i.d.diversity branches ($k = 2$ and $m = 5$).

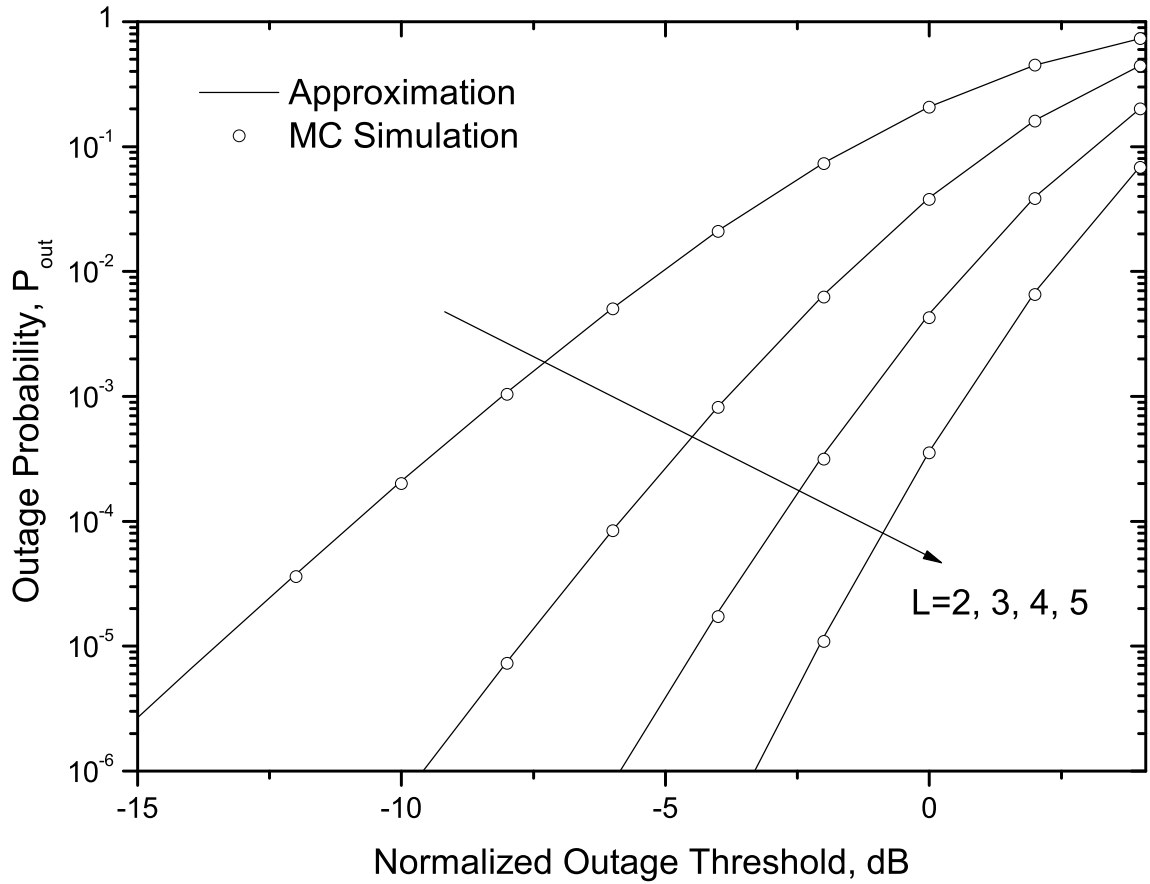

Fig. 2. Comparison of approximate Outage probability and MC simulation results of MRC receivers with i.i.d. diversity branches ($k = 2$ and $m = 5$).

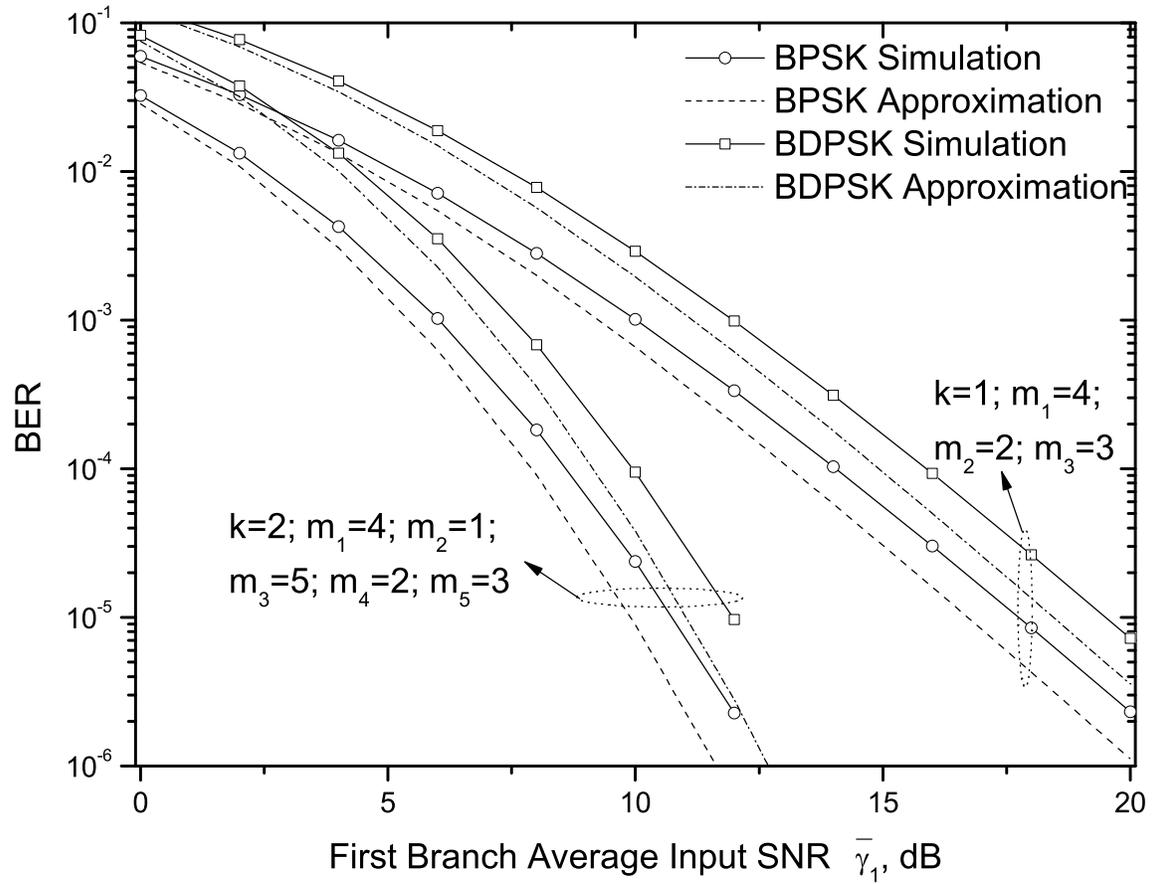

Fig. 3. Comparison of approximate average BER and MC simulation results of MRC receivers with i.n.i.d. diversity branches ($\delta = 0.5$).

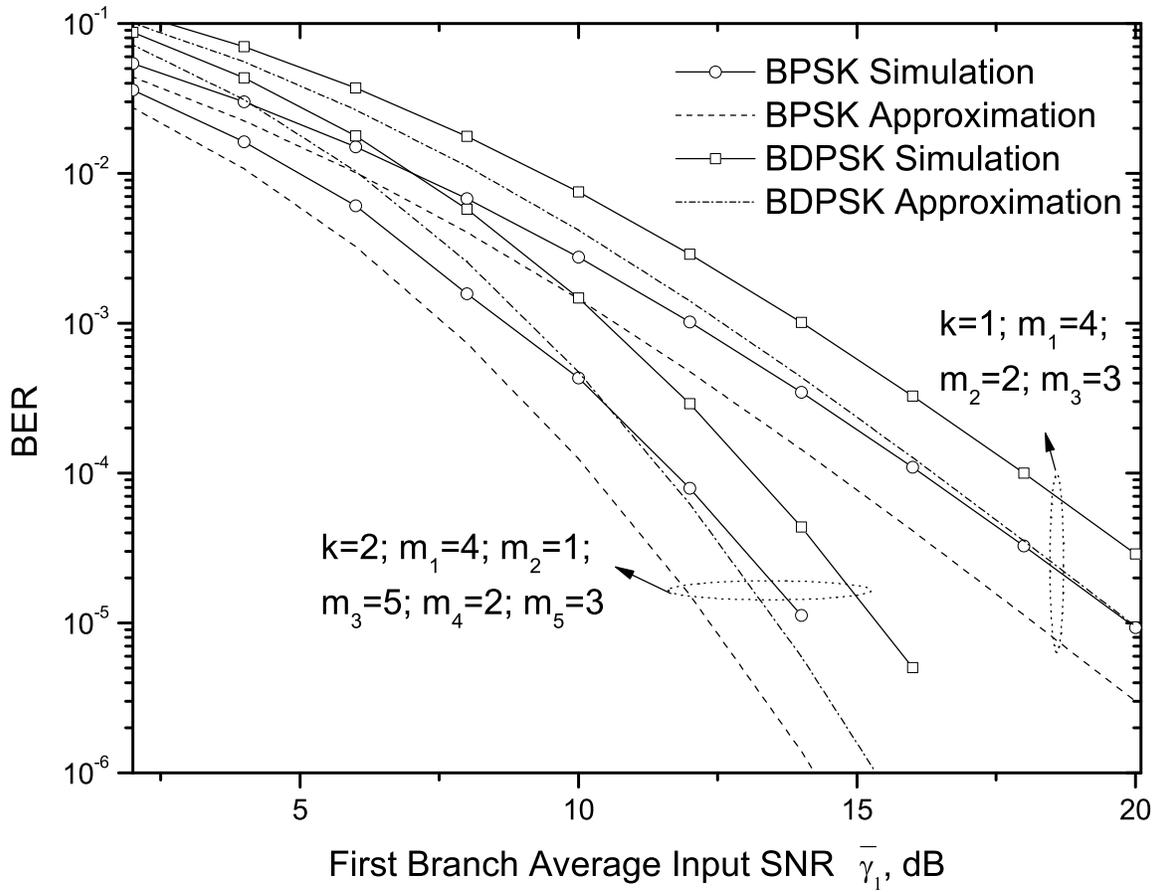

Fig. 4. Comparison of approximate average BER and MC simulation results of MRC receivers with i.n.i.d.diversity branches ($\delta = 1$).

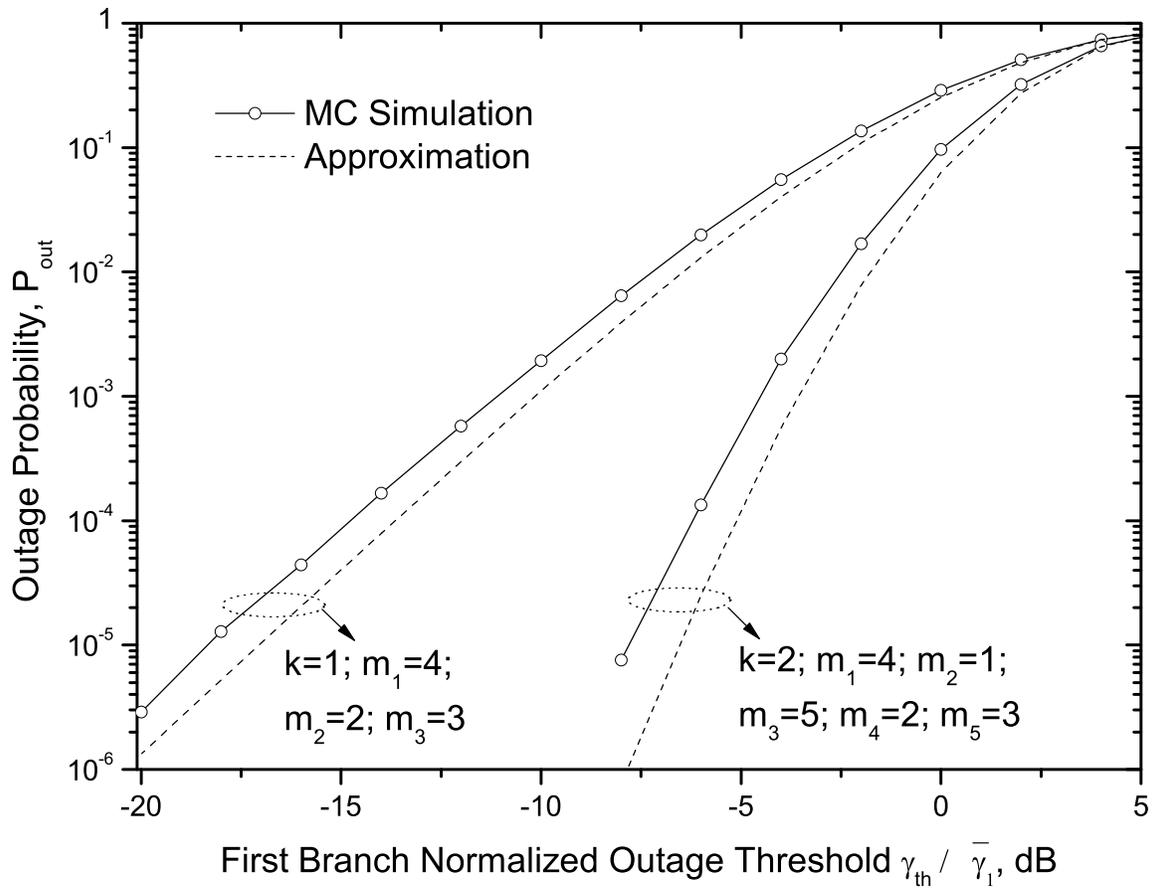

Fig. 5. Comparison of approximate Outage probability and MC simulation results of MRC receivers with i.n.i.d. diversity branches ($\delta = 0.5$).

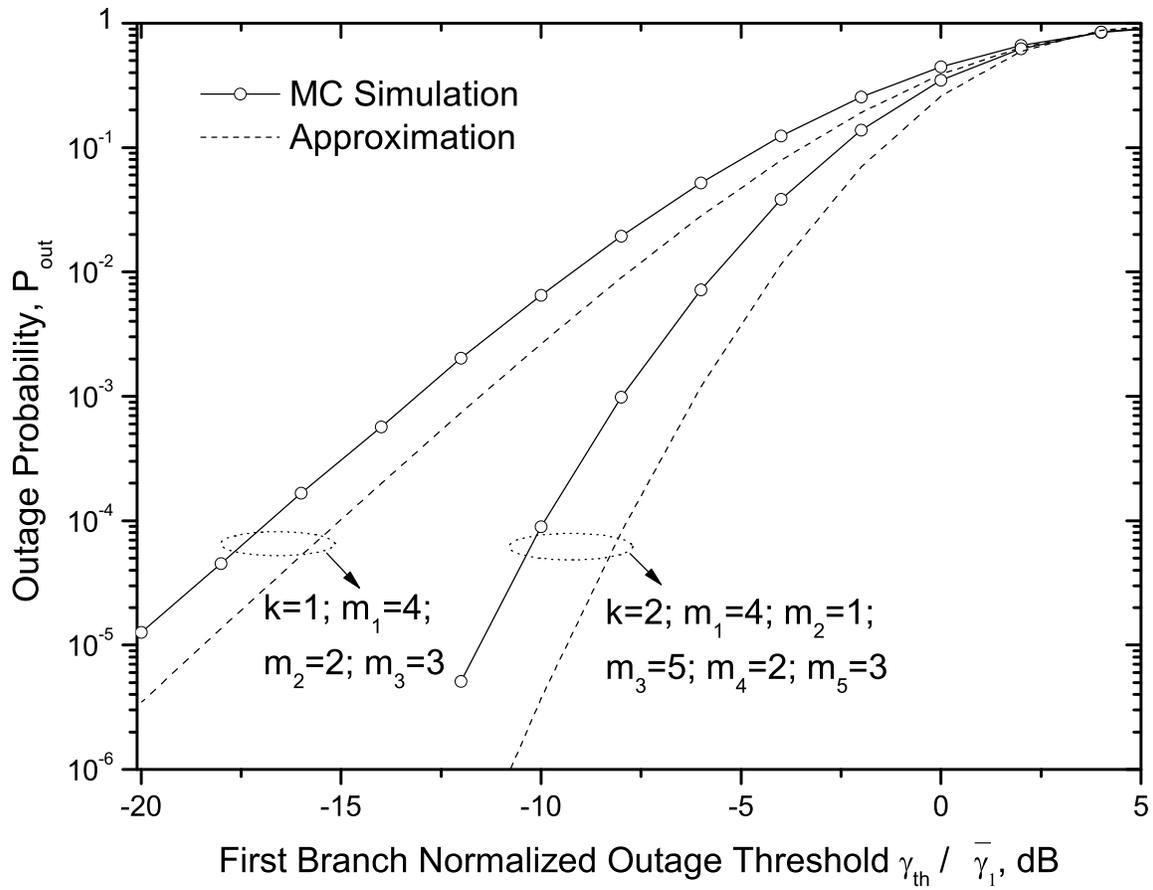

Fig. 6. Comparison of approximate Outage probability and MC simulation results of MRC receivers with i.n.i.d. diversity branches ($\delta = 1$).

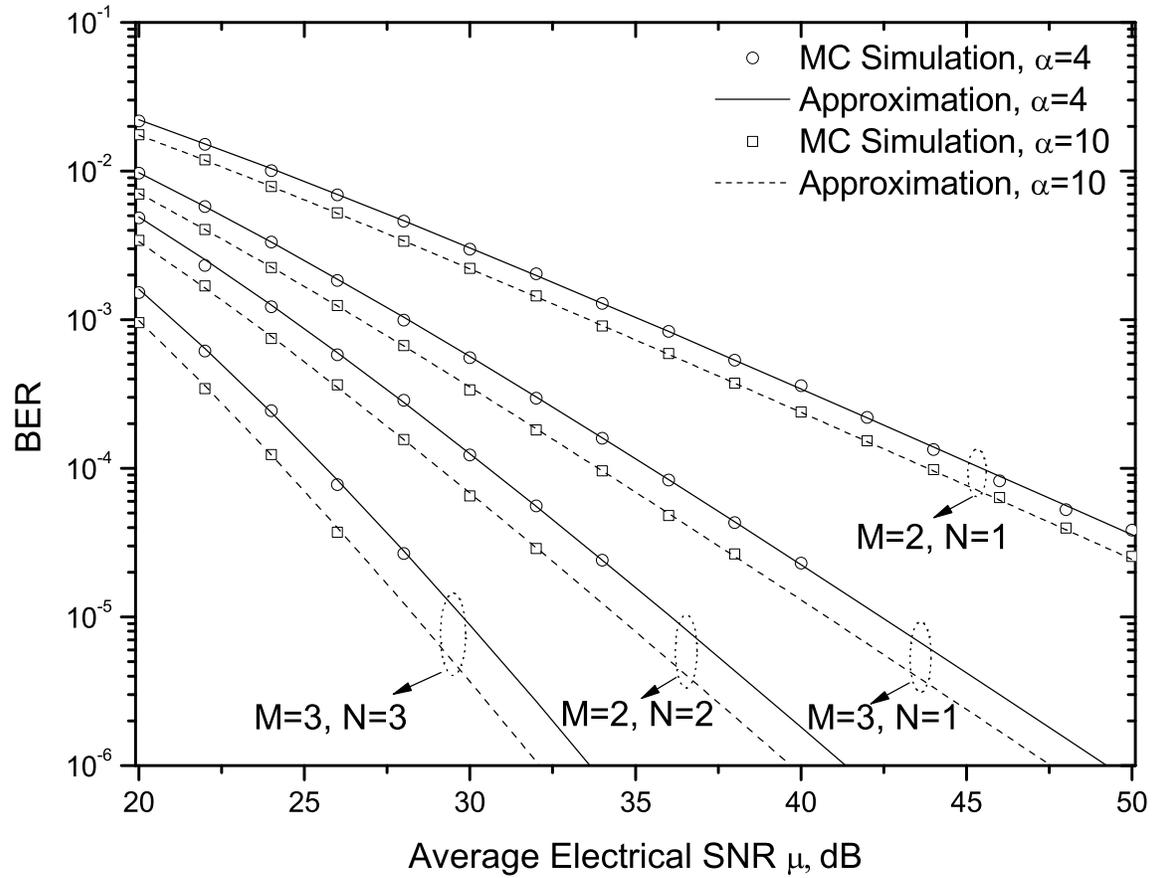

Fig. 7. Comparison of approximate average BER and MC simulation results for various MIMO OW systems over i.i.d. links.

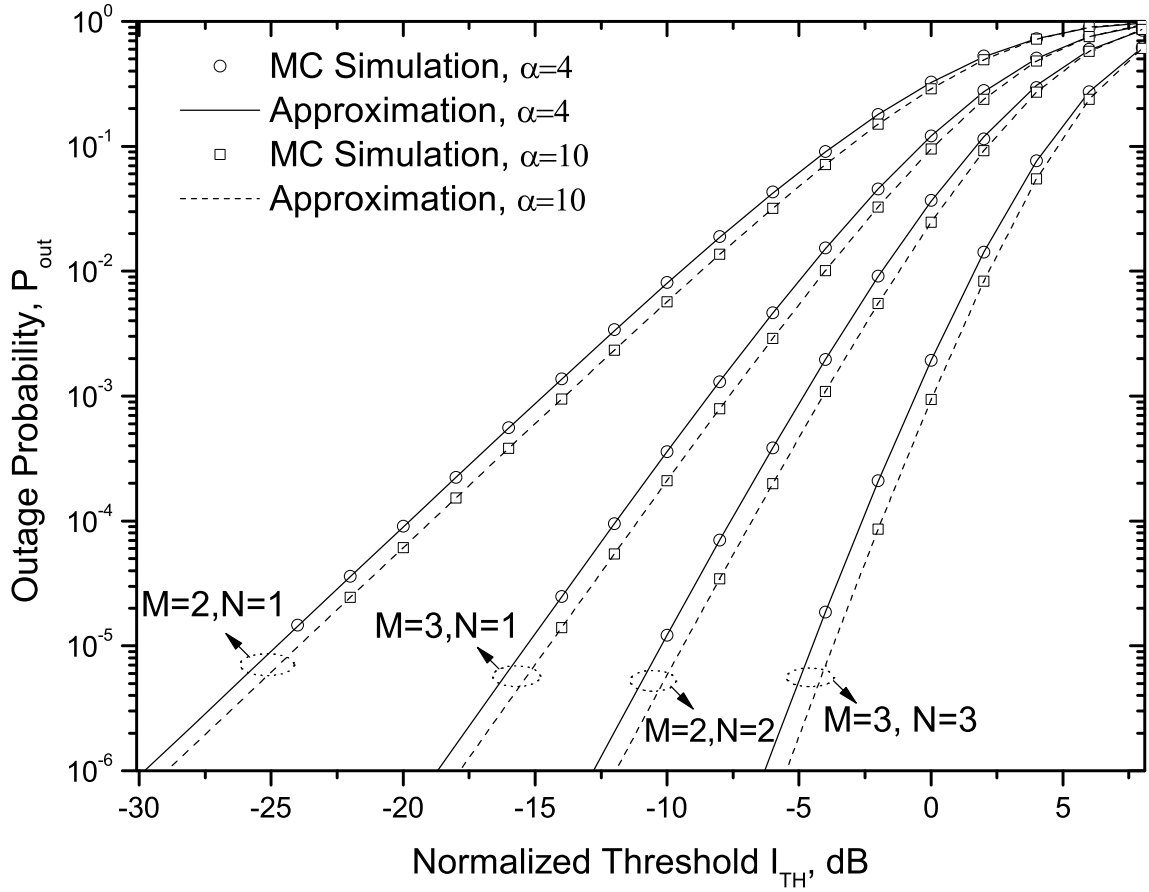

Fig. 8.  Comparison of approximate outage probability and MC simulation results for various MIMO OW systems over i.i.d. links.

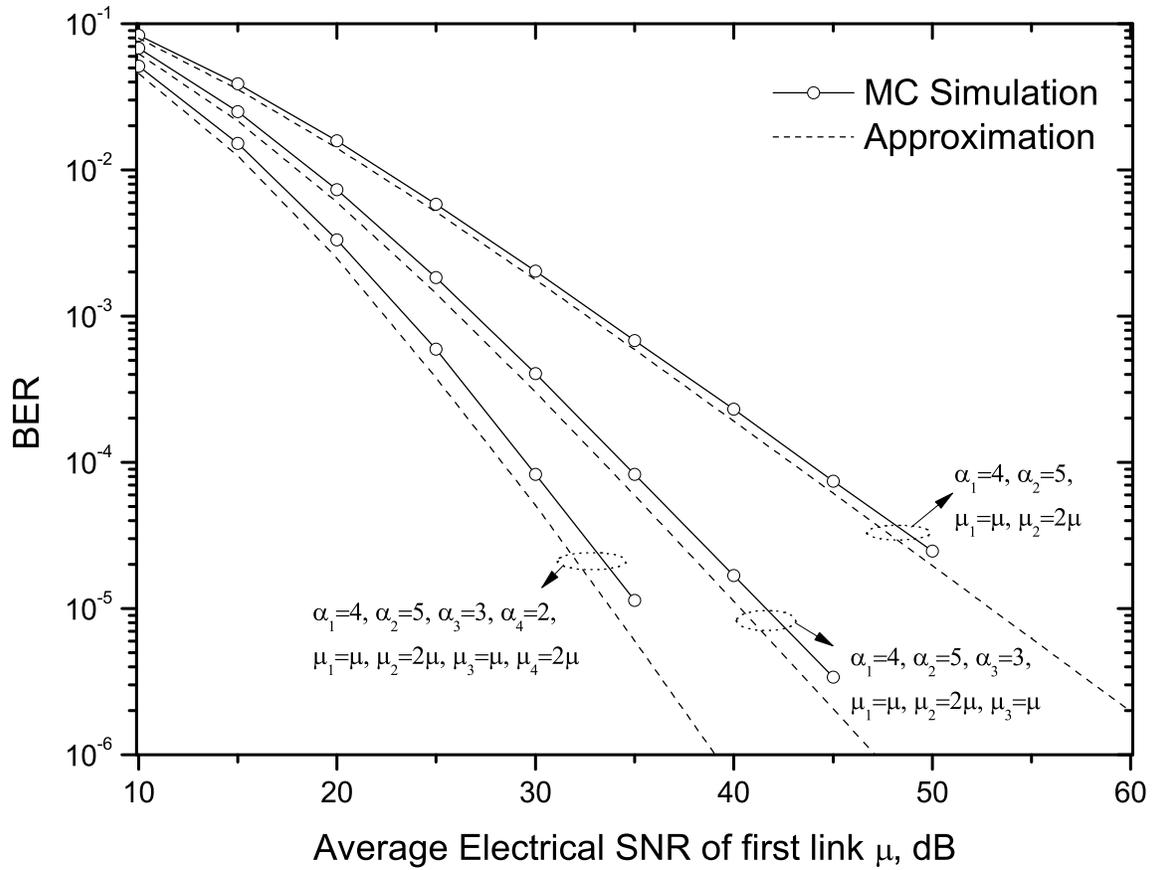

Fig. 9.  Comparison of approximate BER performance and MC simulation results of MIMO systems with $M = 2$ and $N = 1$, $M = 3$ and $N = 1$, and $M = 2$ and $N = 2$ transmit and receive apertures.

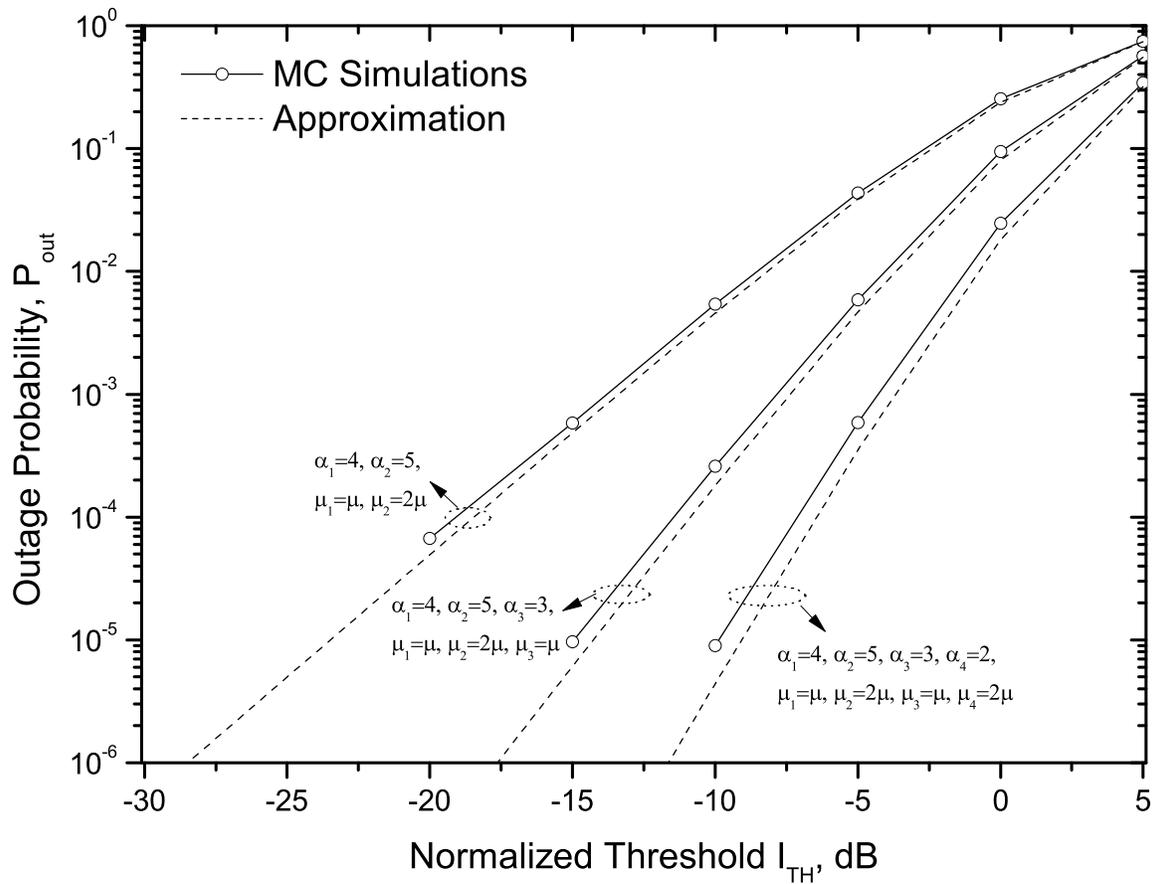

Fig. 10. Comparison of approximate outage probability and MC simulation results of MIMO systems with $M = 2$ and $N = 1$, $M = 3$ and $N = 1$, and $M = 2$ and $N = 2$ transmit and receive apertures.